\documentstyle[11pt,IAUS212,twoside]{article}

\def\edcomment#1{\iffalse\marginpar{\raggedright\sl#1\/}\else\relax\fi}
\reversemarginpar

\begin{document}
\vspace*{1cm}
\title{LBV (Candidate) Nebulae: Bipolarity and Outflows}
 \author{Kerstin Weis\footnotemark}
\footnotetext{Feodor-Lynen-fellow, Alexander-von-Humboldt foundation}
\affil{Max-Planck-Institut f\"ur Radioastronomie, Auf dem H\"ugel 69, 53121
  Bonn, Germany and University of Minnesota, 116 Church Street SE,
  Minneapolis, MN 55455, USA }

\begin{abstract}
The most massive evolved stars (above 50\,M$_{\sun}$) undergo a 
phase of extreme mass loss in which their evolution is reversed 
from a redward to a blueward motion in the HRD. In this phase 
the stars are known as Luminous Blue Variables (LBVs) and they 
are located in the HRD close to the Humphreys-Davidson limit. It 
is far from understood what causes the strong mass loss or what 
triggers the so-called giant eruptions, active events in which 
in a short time a large amount of mass is ejected. Here I will 
present results from a larger project devoted to better understand 
LBVs through studying the LBV nebulae. These nebulae are formed 
as a consequence of the strong mass loss. The analysis concentrates 
on the morphology and kinematics of these nebulae. Of special 
concern was the frequently observed bipolar nature of the LBV 
nebulae. Bipolarity seems to be a general feature and strongly 
constrains models of the LBV phase and especially of the formation
of the nebulae. In addition we found outflows from LBV
nebulae, the first evidence for ongoing instabilities in the nebulae.  
\end{abstract}

\section{LBV nebulae---their morphology and kinematics}

\begin{table}
\begin{center}
\begin{tabular}{cccccc}
\hline
\hline
LBV & host galaxy & size & $v_{\rm exp}$ & morphology \\
(candidate) & & [pc] & [km/s] & \\
\hline
\hline
$\eta$ Carinae & Milky Way & 0.2/0.67 & 600/$10-2000$ & bipolar \\
HR Carinae & Milky Way & 1.3\,$\times$\,0.65 & $75-150$ & bipolar \\
P Cygni & Milky Way & 0.2/0.8 & $110-140$/185 & spherical/clumpy  \\
AG Carinae  & Milky Way & 0.87\,$\times$\,1.16 & 70 & bipolar (?) \\
WRA 751 & Milky Way & 0.5 & 26 & spherical/bipolar \\
He 3-519 & Milky Way & 2.1 & 61 & spherical (?) \\
HD 168625 & Milky Way & 0.13\,$\times$\,0.17& 40 & spherical/clumpy \\
Pistol Star & Milky Way & 0.8\,$\times$\,1.2 & 60 &  spherical/clumpy\\
R127 & LMC & 1.3 & 32 &  spherical/bipolar \\
R143 & LMC & 1.2 & 24 (?) & clumps \\
S61 & LMC & 0.82 & 27 & spherical/outflow (?)\\
S119 & LMC (?) & 1.8 & 26 &  spherical/outflow\\
\hline
Sher 25 (?)& Milky Way & 1 & 70 & bipolar \\
Sk$-69\deg$ 279 (?)& LMC & 4.5 & 14 & spherical/outflow\\
\hline
\end {tabular}
\begin{tabular}{l}
\hline
\hline
\end{tabular}
\end{center}
\end{table}
To better understand why massive stars in the LBV phase 
increase their mass loss  and  
what triggers the sporadic and powerful giant eruptions 
we studied the LBV nebulae, obvious relics
of these processes. Using high-resolution images
and high-dispersion echelle spectra we concentrate on the morphology and
kinematics of all LBV nebulae which are  resolvable  
by HST. This limits our sample to nebulae in the Milkyway, LMC and SMC. 
Due to the page limiting see the following references for 
a detailed description of the
results and individual objects  Weis et al.\ (1997), Weis (2000), 
Weis (2001a,b), Weis \& Duschl (2002), Weis et al.\ (2002, sub.), Weis (2002, 
in prep.). More general info on LBVs and LBV nebulae are published by
Humphreys \& Davidson (1994), Nota et al.\ (1995), Weis (2001b)
and references therein. Here I summarize our global 
results (Weis 2001b, for references on Table 1), 
sizes and expansion velocities of LBV nebulae are given in Tab.\ 1.
Among LBV nebulae $\eta$ Carinae seems special. It 
shows the highest expansion velocities  2000\,km/s  (see Weis et al., this 
proceeding) found in LBV nebulae and its shape is highly bipolar. 
The expansion velocity is unique (Tab. 1) but the morphology not! HR Carinae 
(Weis et al.\ 1997) is bigger in size but with the same bipolar structure.
At the same time the expansion velocity is lower, indicating that 
HR Car might be an older ($\sim$ $4000-9000$ yrs), slowed down 
version of $\eta$ Car's nebula.
Other nebulae  (e.g. WRA 751, R\,127) are not as strong bipolar 
but still show  components which move bi-directional.  
Finally some nebulae are spherical, but reveal a very 
prominent outflow feature (e.g.\ S\,119, Sk$-69\,\deg$279 a new
LBV candidate).
In summary we find that instabilities in LBV nebulae are present and 
that the bipolarity occurs frequently but with different strength.
At this stage we can only speculate whether or not it might be 
due to stellar rotation, asymmetrical/bipolar winds or a binary---or 
something we have not thought of yet. Nevertheless a new constraint to
the evolution of LBVs is that bipolarity has to 
be achieved while forming these nebulae.

\end{document}